\documentclass[final,runningheads]{llncs}
\pdfoutput=1
\usepackage[utf8]{inputenc}
\usepackage[T1]{fontenc}
\usepackage{graphicx}
\usepackage[dvipsnames,table]{xcolor}
\usepackage[final,hidelinks]{hyperref}
\definecolor{fmlinks}{RGB}{48,92,214}
\hypersetup{
	colorlinks,
	linkcolor=black,
	citecolor=black,
	urlcolor=fmlinks
}
\usepackage{comment}
\usepackage{multicol}
\usepackage{amssymb}%
\usepackage{color}

\usepackage{xparse}
\usepackage{expl3}
\usepackage{amsmath}%
\usepackage[nameinlink,capitalize]{cleveref}%
\usepackage[abbreviations]{foreign}
\usepackage{cite}
\usepackage{enumitem}
\usepackage{multirow}
\usepackage{jolie}
\usepackage{orcidlink}%
\def\orcidID#1{\unskip$^{\orcidlink{#1}}$}
\usepackage{tikz}%
\usetikzlibrary{arrows.meta,backgrounds,calc,decorations.markings,fit,positioning,shapes,shapes.geometric,shapes.multipart,shapes.symbols,chains}

\let\concept\emph
\let\conceptname\texttt
\let\pattern\textsc
\newlength{\starheight}
\newcommand{\rank}[1]{%
  \begin{tikzpicture}[baseline]
  \foreach \i in {1,...,3} {
    \pgfmathparse{\i<=#1 ? "fill" : "draw"}
    \edef\content{\pgfmathresult}
    \node[star,draw,\content,star point ratio=2.3,anchor=outer point 3,inner sep=0,minimum width=\starheight] at (\i*\starheight*1.25,0) {};
  }
  \end{tikzpicture}%
}
\newcommand*{\letterrank}[2]{%
  \begin{tikzpicture}[baseline]
  \node[inner sep=0pt,outer sep=0pt,minimum width=\starheight,%
  anchor=base,font=\sffamily,align=center] at (\starheight*0.25,0) {#1};
  \foreach \i in {1,...,3} {
    \pgfmathparse{\i<=#2 ? "fill" : "draw"}
    \edef\content{\pgfmathresult}
    \node[star,draw,\content,star point ratio=2.3,anchor=outer point 3,inner sep=0pt,minimum width=\starheight] at (\i*\starheight*1.25,0) {};
  }
  \end{tikzpicture}%
}
\newcommand{\lE}{{\sffamily E}}
\newcommand{\lM}{{\sffamily M}}
\newcommand{\lI}{{\sffamily I}}
\newcommand{\EMI}{{\sffamily EMI}}
\newcommand{\E}[1]{\letterrank{E}{#1}}%
\newcommand{\M}[1]{\letterrank{M}{#1}}%
\newcommand{\I}[1]{\letterrank{I}{#1}}%
\newcommand{\cellranking}[3]{{%
  \def\arraystretch{1}%
  \vbox to 3.5em {\vfil
  \hbox to 6em {\begin{tabular}{l}
    \E{#1}\\\M{#2}\\\I{#3}
  \end{tabular}}\vfil}}}
\newcommand{\apipattern}[1]{{\scshape #1}}

\newenvironment{recipe}{%
  \par\noindent\rule{\textwidth}{1pt}\par\nopagebreak\noindent
}{%
  \par\noindent\rule{\textwidth}{1pt}\par\nopagebreak
}

\newlength\psl
\def\ps{\the\psl}
\def\cs{2.5pt}
\def\op{{Circle[fill=white,length=\ps]}}
\def\ip{{Circle[fill=black,length=\ps]}}
\def\oc{{Arc Barb[reversed,length=\cs]}}
\def\ic{{Arc Barb[length=\cs]}}
\tikzset{ %
	ms/.style={draw=black,
    outer sep={\dimexpr0.5\psl\relax},
		regular polygon,regular polygon sides=6,
    align=center,#1},
	api *--o/.style={\ip-\op,shorten >=-\ps,shorten <=-\ps},
	api o--*/.style={\op-\ip,shorten >=-\ps,shorten <=-\ps},
	api *--(/.style={\ip-\oc,shorten <=-\ps},
	api o--(/.style={\op-\oc,shorten <=-\ps},
	api o--)/.style={\op-\ic,shorten <=-\ps},
	api )--*/.style={\oc-\ip,shorten >=-\ps},
	api (--o/.style={\ic-\op,shorten >=-\ps},
	ms lbl/.style={align=center,font=\footnotesize,#1},
  database/.style={draw,
    cylinder,
    shape border rotate=90,
    aspect=0.15,
    outer sep={\dimexpr0.5\psl\relax}},
  participants/.style={every node/.style={draw,ms,thin},node distance=0.6cm},
  embedded/.style={double},
  links/.style={
    every node/.append style={
      draw,circle,midway,above,
      inner sep=1pt,outer sep=2pt,font=\scriptsize}},
  invisible nodes/.style = {
    every node/.append style={draw=none,text opacity=0}},
  context/.style={
    draw,black,rectangle,thick,densely dotted,rounded corners,
    inner xsep=-\cs,inner xsep=3, inner ysep=1,outer sep=3},
  context lbl/.style={font=\tiny,black},
  margin/.style={fit=(current bounding box),inner ysep=2pt,inner xsep=2pt}
}

\makeatletter
\newcommand{\customlabel}[4][0]{%
  \protected@write\@auxout{}{\string\newlabel{#3}{{#4}{\thepage}{#4}{#3}{}}}%
  \protected@write\@auxout{}{\string\newlabel{#3@cref}{{[#2][#1][#1]#4}{\thepage}{}{}{}}}%
}
\makeatother
\Crefname{lstlisting}{Listing}{Listings}
\def\llabel#1{%
  \customlabel{line}{#1}{\thelstnumber}%
  \hypertarget{#1}{}%
}

\begin{document}
\title{A Conceptual Framework for API Refactoring in Enterprise Application Architectures}
\titlerunning{A Conceptual Framework for API Refactoring}
\author{Fabrizio Montesi\inst{1}\orcidID{0000-0003-4666-901X} \and
  Marco Peressotti\inst{1}\orcidID{0000-0002-0243-0480} \and
  Valentino Picotti\inst{1}\orcidID{0000-0001-7713-1461} \and
  Olaf Zimmermann\inst{2}}
\authorrunning{F.~Montesi et al.}
\institute{Department of Mathematics and Computer Science, University of Southern Denmark
  \email{\{fmontesi,peressotti,picotti\}@imada.sdu.dk} \and
  University of Applied Sciences of Eastern Switzerland
  \email{olaf.zimmermann@ost.ch}}

\maketitle              %
\begin{abstract}
Enterprise applications are often built as service-oriented architectures, where the individual services are designed to perform specific functions and interact with each other by means of well-defined APIs (Application Programming Interfaces).
The architecture of an enterprise application evolves over time, in order to adapt to changing business requirements.
This evolution might require changes to the APIs offered by services, which can be achieved through appropriate API refactorings.

Previous studies on API refactoring focused on the effects on API definitions, with general considerations on related forces and smells.
So far, instead, the development strategy for realising these refactorings has received little attention.
This paper addresses exactly this aspect.

We introduce a conceptual framework for the implementation of API refactorings. Our framework elicits that there are important trade-offs and choices, which significantly affect the efficiency, maintainability, and isolation properties of the resulting architecture.
We validate our framework by implementing several refactorings that introduce established API patterns with different choices, which illustrates the guiding principles offered by our framework.
Our work also elicits, for the first time, how certain programming language features can reduce friction in applying API refactoring and open up more architectural choices.

\end{abstract}

\section{Introduction}
\label{sec:intro}

Enterprise applications are distributed information systems designed to support the needs of complex organisations, for example by managing business processes and handling data~\cite{F12}.
Such applications are often built using (micro)service-oriented architectures, where individual services perform specific functions and interact through well-defined APIs (Application Programming Interfaces)~\cite{DGLMMMS17}.

As organisations evolve, so do their applications: enterprise application architectures must continuously adapt to changing business requirements~\cite{EP15,FPK17,EPWS21}.
In the context of APIs, this prompted studying \emph{API refactoring}: the modification of interfaces to improve quality attributes, such as efficiency~\cite{SZ21,SZ23}.

Recently, a catalogue of API patterns provided a basis for API refactorings~\cite{ZSLZP2022,SZ23}.
Previous studies mainly focused on the high-level forces and smells, like modifiability and high latency, that motivate and guide API refactorings.
Conversely, it is yet unknown how developers are supposed to implement an API refactoring and assess its quality; 
a research gap that we address in this work.

In this paper, we introduce \EMI{} (efficiency, maintainability and isolation), a conceptual framework for assessing the implementation of API refactoring in service architectures.
\EMI{} centres around two dimensions: 
1) \emph{generality}, which assesses the degree of abstraction of the refactored API source code; %
2) \emph{distribution}, which elicits where the refactored API source code resides.
Realising the combination of both dimensions results in six development strategies, 
each representing design choices for the implementation with respective trade-offs.
The trade-offs pertain the quality aspects of \emph{efficiency} (\lE{}), \emph{maintainability} (\lM{}), and \emph{isolation} (\lI) of the resulting architecture.
We score each of these three aspects from $1$ to $3$ for our strategies, yielding the \EMI{} score for API refactoring.
Clearly, there is no silver bullet: no strategy scores perfectly ($9$), emphasising the importance of making conscious implementation decisions.

To validate our framework in practice, we apply it to several API refactorings implemented in the service-oriented programming language Jolie~\cite{MGZ14}. %
We choose Jolie because of its abstraction and expressiveness capabilities.
Firstly, Jolie offers a technology-agnostic language for defining APIs
and native constructs for declaring endpoints that consume and provide APIs. This makes our refactorings direct and easy to understand.
Secondly, Jolie offers features not present in other programming languages, such as API polymorphism and aggregation (the merging of endpoints), allowing us to implement design choices declaratively.

In more detail, we apply our six development strategies to the same refactoring: the introduction of the \apipattern{API Key} pattern -- which rejects requests without a valid key (\cref{sec:jolie}) -- to a service that offers a catalogue of scientific publications.
We then broaden our study to patterns that do not require behavioural changes, \apipattern{Merge Endpoints} and \apipattern{Version Identifier}, reaching modular solutions (\cref{sec:other-patterns}).
A main finding is that our framework can be used to distill systematic recipes for API refactoring, which developers can mechanically apply step by step to achieve an implementation with a declared \EMI{} score (\cref{sec:recipes}).

In summary, 
we contribute the \EMI{} framework, a scheme to assess the adaption of API refactorings, 
we validate \EMI{} by applying API refactorings to (micro)service architectures, and provide canonical recipes to obtain certain \EMI{} scores for the architecture evolution.
Thus, our work offers developers the possibility to assess and implement API refactorings in practice and lays the grounds to explore the impact of refactoring APIs in further detail.

\section{Related Work}
\label{sec:related}

Our study builds on the reference catalogue of patterns for API design~\cite{ZSLZP2022}, which addresses the challenge of remote API design~\cite{ZSLPZ19} through peer-reviewed patterns published in the period 2017--2020~\cite{LZPZS19,ZPLZS20,ZLZPS20,SZZLP18,ZSLZ17}.

API refactoring was previously investigated based on the same catalogue~\cite{SZ21,SZ23}. 
Those studies focus on architectural considerations and especially \emph{why} and \emph{when} an API pattern should be introduced, considering forces and smells.
Differently, our work is the first to investigate \emph{how} an API pattern is implemented. 
Specifically, our interest lies in the different choices that one can make regarding the code of a refactoring, and the quality trade-offs that they yield.
Another difference is that we study how to refactor also the implementation of an API, whereas previous work discusses only how to refactor the API definition.
Our frameworks can be seen as a refinement of Attribute-Driven Design and Architecture Trade-off Analysis Method which are concerned with informing and assessing architectural decisions in light of quality attribute requirements \cite{BCK2012}.

Jolie is an emerging programming language for service-oriented computing~\cite{MGZ14}, which has recently been gaining traction because of its native features for defining and composing services~\cite{BKMRST16,GM19,GMG23,GDM16,GKSMM16,GGLZ18}.
Previous work has validated Jolie in different domains, including system integration~\cite{M16,GGLZ18}, Internet of Things~\cite{GKSMM16,GGLZ18}, journalism~\cite{CKMV23}, and cloud provisioning~\cite{GAV12}.

The work nearest to ours is perhaps the implementation of an API- and location-agnostic circuit breaker~\cite{MW18} -- a pattern for increasing resilience~\cite{N07}.
In that development, Jolie's high-level abstractions are used to obtain a flexible implementation and experiment with different deployment strategies (in the client, in a forwarding proxy, or in the server). These have important security implications~\cite{MW18} that were later confirmed in an independent study~\cite{C19}.
The developments in~\cite{MW18} fall under the Parametric/Adjacent and Parametric/External strategies in our framework (see \cref{sec:strategies}).
Our interest in the present work is much more general: rather than focusing on a specific use case, we formulate a framework that can be used to reason about any API refactoring.
Furthermore, the quality aspects considered here are not considered in~\cite{MW18}.

Many frameworks and languages for service-oriented systems have been proposed, including \href{https://spring.io/projects/spring-boot}{Spring Boot}, \href{https://expressjs.com/}{Express} for Node.js, Ballerina~\cite{O19}, and WS-BPEL~\cite{bpel}.
Some of Jolie's features that we use in our work can be found or implemented in these technologies. Aggregation and redirection -- Jolie for merging and redirecting endpoints -- recall the routing mechanism in Express.
Jolie uses structural typing, like Ballerina~\cite{O19}, which is more suitable to networked (and multi-technology) systems than nominal typing.
On top of offering all the features we need in a single package, Jolie's primitives are designed to make the resulting APIs statically known, which is useful in the context of API design.

\section{The \EMI{} Framework for API Refactoring}
\label{sec:strategies}

\begin{table}[t]
\begin{center}{
\addtolength{\tabcolsep}{1ex}
\def\arraystretch{1.5}
\def\rsep{-1.2ex}
\begin{tabular}{|c|c|c|c|c|}
 \cline{3-5}
  \multicolumn{2}{c|}{}&\multicolumn{3}{c|}{\textbf{Distribution}}\\
  \cline{3-5}
  \multicolumn{2}{c|}{}&Internal&Adjacent&External\\
  \hline
  \multirow{6}{*}{\rotatebox[origin=c]{90}{~~~~~\textbf{Generality}}}
  &          &\E{2}&\E{2}&\E{1}\\[\rsep]
  &Parametric&\M{2}&\M{3}&\M{3}\\[\rsep]
  &          &\I{1}&\I{2}&\I{3}\\
  \cline{2-5}
  &          &\E{3}&\E{2}&\E{1}\\[\rsep]
  &Ad-hoc    &\M{1}&\M{2}&\M{2}\\[\rsep]
  &          &\I{1}&\I{2}&\I{3}\\
  \hline
\end{tabular}}
\end{center}

\caption{The \EMI{} framework.}
\label{tab:emi}
\end{table}

We now present our conceptual framework for API refactoring -- the \EMI{} framework.
It is depicted in \cref{tab:emi}. We explain it in the rest of this section.

API refactoring changes both an interface and its implementation, while improving at least one quality attribute~\cite{SZ21,SZ23}.
This may affect the external behaviour of an API observed by clients, without altering its capabilities.

We introduce some terminology.
In the remainder, we refer to the changes introduced by an API refactoring as the \emph{new functionality}, bearing in mind that such functionality does not alter the feature set of the API \cite{SZ23}.
In line with the API domain model of \cite{ZSLZP2022}, we consider an API to be a collection of operations that can be invoked by clients. Services can offer APIs through \emph{endpoints}, which expose operations at a designated location according to a given transport protocol.
We call such services \emph{API providers}.
We distinguish the API and implementation that we start from and then end up with after a refactoring with the prefixes \emph{original} and \emph{refactored}.

\subsection{Generality and distribution}

The \EMI{} framework focuses on two dimensions to assess the quality attributes of the implementation of an API refactoring: \emph{generality} and \emph{distribution}.

The \emph{generality} dimension concerns whether the implementation of the new functionality depends on or abstracts from the definition of the original API. We identify two possibilities.
\begin{description}
\item[Ad-hoc] The code of the new functionality depends on hardcoded information on the names, types, or behaviours of the operations in the original API.
\item[Parametric] The code of the new functionality abstracts from the names, types, or behaviours of the operations in the original API.
\end{description}
Generality serves as an indicator of the logical coupling between the new code and the old. It is significant because API patterns provide, at least conceptually, reusable solutions to recurring problems. Thus, in a way, generality indicates how much this reusability is achieved in real code.

\begin{figure}[t]
\centering
\renewcommand{\arraystretch}{3}
\begin{tabular}{lcr}
  Initial Position&
  \begin{tikzpicture}[anchor=base,baseline]
  \begin{scope}[start chain=1,participants]
  \node[on chain=1] (0) {C};
  \node[on chain=1] (1) {O};
  \node[fit=(1),context] {};
  \end{scope}
  \begin{scope}[links]
  \draw[api o--*] (0) -- (1);
  \end{scope}
  \node[margin] {};
  \end{tikzpicture}
  &
  \multirow{4}{*}[-.8em]{
  \renewcommand{\arraystretch}{1.2}
  \begin{tabular}{cl}
    \tikz[anchor=base,baseline]{\node[ms,inner sep=1.7] (0,0) {C};\node[margin]{};}&Client\\
    \tikz[anchor=base,baseline]{\node[ms,inner sep=1.7] (0,0) {O};\node[margin]{};}&Original API provider\\
    \tikz[anchor=base,baseline]{\node[ms,inner sep=1.7] (0,0) {R};\node[margin]{};}&Refactored API provider\\
    \tikz{\node[circle,draw,fill,inner sep=0,minimum size=\ps] (0,0) {};}&API provider endpoint\\
    \tikz{\node[circle,draw,fill=white,inner sep=0,minimum size=\ps] (0,0) {};}&API consumer endpoint\\
    \tikz{\draw (0,0)--(0.6,0);\node[margin,inner ysep=2.5]{};}&Network link\\
    \tikz{\draw[embedded] (0,0)--(0.6,0);\node[margin,inner ysep=2.5]{};}&In-memory link\\
    \tikz[anchor=base,baseline]{\node[context,inner ysep=6,inner xsep=2]{\phantom{CC}};\node[margin]{};}&Deployment location\\
 \end{tabular}
  }\\\cline{1-2}
  Internal&
  \begin{tikzpicture}[anchor=base,baseline]
  \begin{scope}[start chain=1,participants]
  \node[on chain=1] (0) {C};
  \node[on chain=1] (1) {R};
  \node[fit=(1),context] {};
  \end{scope}
  \begin{scope}[links]
  \draw[api o--*] (0) -- (1);
  \end{scope}
  \node[margin] {};
  \end{tikzpicture}
  \\\cline{1-2}
  Adjacent&
  \begin{tikzpicture}[anchor=base,baseline]
  \begin{scope}[start chain=1,participants]
  \node[on chain=1] (0) {C};
  \node[on chain=1] (1) {R};
  \node[on chain=1] (2) {O};
  \end{scope}
  \begin{scope}[links]
  \draw[api o--*] (0) -- (1);
  \draw[api o--*,embedded] (1) -- (2);
  \end{scope}
  \node[fit=(1) (2),context] (context12) {};
  \node[margin] {};
  \end{tikzpicture}
  &\\\cline{1-2}
  External&
  \begin{tikzpicture}[anchor=base,baseline]
  \begin{scope}[start chain=1,participants]
  \node[on chain=1] (0) {C};
  \node[on chain=1] (1) {R};
  \node[on chain=1] (2) {O};
  \end{scope}
  \begin{scope}[links]
  \draw[api o--*] (0) -- (1);
  \draw[api o--*] (1) -- (2);
  \end{scope}
  \node[fit=(1),context] (context1) {};
  \node[fit=(2),context] (context2) {};
  \node[margin] {};
  \end{tikzpicture}
  &\\
\end{tabular}
\caption{Possible choices for distribution.}
\label{fig:distribution}
\end{figure}
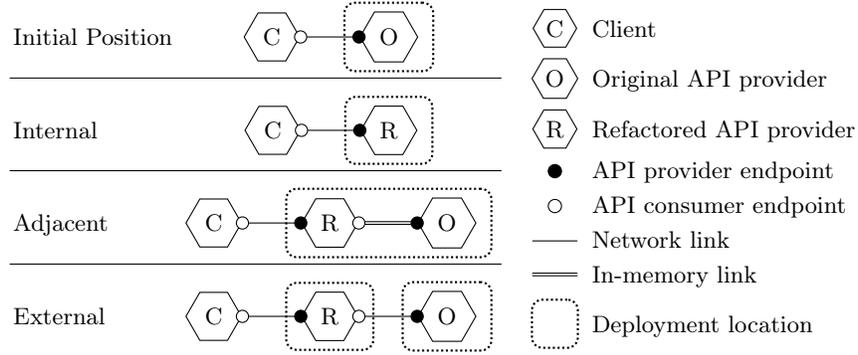

The \emph{distribution} dimension concerns where the code for the new functionality is located in relation to the original API provider and its clients.
There are three possibilities, depicted in \cref{fig:distribution}.
\begin{description}
\item[Internal] The code of the new functionality is mixed with the code of the original API provider. Thus, they share state. After the refactoring, the original API provider becomes the refactored API provider.
\item[Adjacent] The refactored API provider is a separate service. They have separate state and are executed independently, but they are deployed such that they can communicate efficiently through local resources (local memory channels, inter-process communication, loopback network interfaces, etc.).
\item[External] The refactored API provider is a separate service.
It is deployed remotely from the original, and thus can communicate with it only through network communication.
\end{description}

\subsection{\EMI{} scores}

The combination of the axes of generality and distribution gives rise to six possible development strategies, each presenting different trade-offs.
To help in navigating these trade-offs, we score each strategy on three quality attributes using a three-level scale (\rank1, \rank2, or \rank3): efficiency (\lE), maintainability (\lM), and isolation (\lI).
We explain each score next.

\begin{description}
\item[Efficiency (\lE)]
\begin{description}
\item%
\item[\E3] The new functionality is implemented optimally, with no extra overhead caused by design choices.
\item[\E2] Design choices cause extra overhead in terms of local resources (memory, local communication, etc.).
\item[\E1] Design choices cause extra overhead in terms of remote resources (e.g., network communication).
\end{description}
\item[Maintainability (\lM)]%
\begin{description}
\item%
\item[\M3] The original and refactored API providers can be maintained independently.
\item[\M2] The implementations of the new functionality and the original API provider are separate but tightly coupled.
\item[\M1] The implementations of the new functionality and the original API provider are completely mixed.
\end{description}
\item[Isolation (\lI)]%
\begin{description}
\item%
\item[\I3] The original and refactored API providers do not share any local resources for their execution.
\item[\I2] The original and refactored API providers share execution resources (e.g., CPUs, memory), but do not share state and interact purely by means of the original API.
\item[\I1] The new functionality and the original API implementation share internal program state (e.g., stack, variables, heap).
\end{description}
\end{description}

The levels of these scales are intentionally broad, in order to avoid being tied up by very specific technological details. This is in line with the technology-agnosticism of microservices~\cite{DGLMMMS17}.

\subsection{Scoring development strategies}

We end the presentation of our framework with an analysis that justifies the \EMI{} scores of each development strategy, referring also to examples of API patterns and technologies where relevant.

\begin{description}
\item[Ad-hoc/Internal (\E3 \M1 \I1)] This is the most efficient strategy, because the new functionality is implemented directly by changing the behaviour of the original API provider.
Thus, the code of the new functionality encounters no unnecessary overhead in integrating with the original implementation.
For example, introducing the \apipattern{Pagination} pattern to an implementation that queries a database gives the possibility to modify the query in order to retrieve fewer results -- those for the page being requested.
For the very same reasons, however, this is also the least maintainable and isolated choice, since the new code is mixed and shares all resources with the old code. Examples of this strategy are shown in \cref{sec:jolie,sec:recipes}.
\item[Parametric/Internal (\E2 \M2 \I1)] %
This strategy trades some efficiency for maintainability by abstracting from the operation names and behaviours of the original API. The code of the new functionality can be reused across different APIs, but has limited access to changing their behaviour: the new functionality can only intercept, modify, and conditionally forward request and response message to and from the original implementation.
Examples of this strategy are implementations adopting \href{https://jakarta.ee/specifications/platform/10/apidocs/jakarta/servlet/filter}{Java Servelet Filters} or \href{https://expressjs.com/en/guide/using-middleware.html}{Express middleware functions}.
\item[Ad-hoc/Adjacent (\E2 \M2 \I2)] 
Compared to Ad-hoc/Internal, implementing the new functionality in a separate component trades some efficiency for partially improved maintainability and isolation.
However, the new functionality remains coupled with the original API (ad-hoc), so changes to the original API require updating the refactored API provider, too. Thus, maintainability is still not ideal.
Improved isolation comes at the cost of some overhead in the interaction between the refactored and original API providers.
Efficiency is further affected by the new functionality not having access to changing the internal behaviour of the original API provider.
This strategy can be implemented with, for example, the \href{https://learn.microsoft.com/en-us/azure/architecture/patterns/sidecar}{sidecar pattern}, the \href{https://learn.microsoft.com/en-us/azure/architecture/patterns/ambassador}{ambassador pattern}, or Jolie's embedded services (see \cref{sec:jolie}).

\item[Parametric/Adjacent (\E2 \M3 \I2)] This strategy has the same efficiency and isolation characteristics as the previous one, but greatly improved maintainability by decoupling the implementation of the new functionality from the operation names and message types of the original API.
The sidecar and ambassador patterns are again useful to implement this strategy. Jolie's embedded services combined with couriers and interface extenders (see \cref{sec:jolie}) offer an interesting solution, because the refactored API can be automatically and statically computed.

\item[Ad-hoc/External (\E1 \M2 \I3)] The strategy with the highest level of isolation, since the new functionality interacts with the original API provider only via remote access.
For the same reason, it is also the least efficient strategy.
This strategy does not achieve the highest maintainability score due to the coupling between the new functionality and the original API.
This strategy can be implemented simply by developing a proxy service offering the refactored API and forwarding each operation invocation to the original API provider when appropriate.
\item[Parametric/External (\E1 \M3 \I3)] This strategy has the same efficiency and isolation scores as the previous one, but also the highest maintainability score for the same reason given for Parametric/Adjacent.
\end{description}

No strategy scores a perfect $9$. The reason lies in the unavoidable tension between efficiency and isolation: optimal efficiency requires sharing resources, which prevents achieving optimal isolation.

\section{Validation}
\label{sec:validation}

In this section, we validate our framework by applying it in depth -- exploring all our strategies for a single pattern -- and in breadth -- applying selected strategies to other patterns.

\subsection{API Key in Jolie}
\label{sec:jolie}

We illustrate the use of our framework by applying each strategy to a concrete use case: the introduction of the \apipattern{API Key} pattern to a service managing a catalogue of scientific publications.
\apipattern{API Key} identifies clients through respective unique keys, which must be included in requests.

We code our examples in Jolie.
In Jolie, the operations and message types of an API are defined as an \jo{interface}.
The next interface defines the API of our publication catalogue service.
\begin{jolielisting}
type Publications: { publications*: Publication }
type Publication: Proceeding | InProceeding | Article

interface PubCatInterface {
  RequestResponse:
    getAuthorPubs( {authorId: string} )( Publications )
    getConfPubs( {confId: string} )( Publications )
}
\end{jolielisting}
\jo{PubCatInterface} comprises two operations: \jo{getAuthorPubs}, which expects the unique identifier of an author (as the field \jo{authorId} of the request message) and returns all their publications (message type \jo{Publications}); and \jo{getConfPubs}, which given a conference identifier (\jo{confId}) returns the  publications of that conference.
The type \jo{Publications} describes a record with a field \jo{publications} containing zero o more (\jo{*}) values of type \jo{Publication}. 
\jo{Publication} is the union of three types (omitted) corresponding to proceedings (\jo{Proceeding}), papers in proceedings (\jo{InProceeding}), and journal articles (\jo{Article}).

Interfaces are offered to clients by defining an \jo{inputPort}, Jolie for an endpoint that accepts remote invocations. An input port is defined inside of its enclosing \jo{service} and commits to a concrete \jo{location} and transport \jo{protocol} (HTTP, SOAP, binary protocols, etc.).
The definition of our publication catalogue service is given next (abstracting some internal implementation details).
\begin{jolielisting}[][caption={Original API Provider.},label=lst:o:service]
/* Service definition */
service PubCat {
  /* API Endpoint */
  inputPort ip { (*\llabel{line:o:inputport-b}*)
    location: "socket://localhost:8080"(*\llabel{line:o:location}*)
    protocol: http { format = "json" }
    interfaces: PubCatInterface
  } (*\llabel{line:o:inputport-e}*)
  /* Behaviour */
  main { (*\llabel{line:o:main-b}*)
    [ getAuthorPubs( request )( response ) { /* fetch the data from db */ } ]
    [ getConfPubs( request )( response ) { /* fetch the data from db */ } ]
  } (*\llabel{line:o:main-e}*)
}
\end{jolielisting}
In \crefrange*{line:o:inputport-b}{line:o:inputport-e}, service \jo{PubCat} exposes \jo{PubCatInterface} on TCP port $8080$ over the HTTP protocol with message payloads in JSON format.
Its implementation (\crefrange*{line:o:main-b}{line:o:main-e}) consists of an \emph{input choice} that can react to any invocation of the operations it lists.
Each branch in the choice has the form:
\begin{equation*}
\mbox{\jo![ operation( request )( response )\{ $\; B \;$ \} ]!}
\end{equation*}
where \jo{operation} is the name of the operation, \jo{request} and \jo{response} are the input and output parameters, and $B$ is the code block computing the response.

Introducing the \apipattern{API Key} pattern requires extending request message types with an additional field \jo{apiKey}
(storing the key as a \jo{string})
and declaring a faulty response message \jo{NotAuthorised} for invocations with invalid keys.
The refactored API is given next.
\begin{jolielisting}[][firstline=2,caption={Refactored API.},label=lst:r:api]
type Error: { code: int, message: string }
interface PubCatInterfaceWithAPIKey {
  RequestResponse:
    getAuthorPubs( {authorId: string, apiKey: string} )( Publications )
      throws NotAuthorised
    getConfPubs( {confId: string, apiKey: string} )( Publications )
      throws NotAuthorised
}
\end{jolielisting}

The refactoring of service \jo{PubCat} and its interface \jo{PubCatInterface} to obtain a service exposing the refactored API \jo{PubCatInterfaceWithAPIKey} can be accomplished following any of the strategies outlined in \Cref{sec:strategies}. We illustrate their application and elicit the different features of Jolie that come to aid.

\paragraph{Ad-hoc/Internal} We directly modify the code of both the original interface \jo{PubCatInterface} and the service \jo{PubCat}. \jo{PubCatInterface} becomes the refactored API \jo{PubCatInterfaceWithAPIKey} above. 
In \jo{PubCat}, instead, the implementation of each operation is edited to validate the API key in the request message.
\begin{jolielisting}
service PubCat {
  inputPort ip {... interfaces: PubCatInterfaceWithAPIKey }
  main {
    [ getAuthorPubs( request )( response ) {
      /* check validity of request.apiKey */
      if( isKeyValid ) { /* fetch the data from db */ }
      else { throw NotAuthorised( /* fault data */ ) }
    } ]
    [ getConfPubs( request )( response ){ /* as for getAuthorPubs */ } ]
  }
}
\end{jolielisting}

\paragraph{Ad-hoc/External}
We introduce a new service, \jo{PubCatWithAPIKey}, with an endpoint exposing the interface \jo{PubCatInterfaceWithAPIKey}.
This service acts as an adapter for the original API provider, \jo{PubCat}, which remains unchanged.
The implementation of the \apipattern{API Key} pattern is entirely confined to the new service, which forwards valid invocations to \jo{PubCat}.
This requires the service \jo{PubCatWithAPIKey} to declare an \emph{output port} (\cref{line:ae:outputport}) pointing to the API endpoint of \jo{PubCat}.
Its implementation (\crefrange{line:ae:main-b}{line:ae:main-e}) consists of an input choice where each operation checks the validity of the key in the request (\jo{request.apiKey}).
If the key is valid, then the key is erased from \jo{request} (\cref{line:ae:undef}) before invoking the original operation \jo{getAuthorPubs@pc} to obtain the intended \jo{response}.
Otherwise, the service replies with a faulty \jo{NotAuthorised} message.
Although the implementations of refactored and original API providers are separate, they must be kept in sync wrt future changes to the API, resulting in a negative impact to maintainability.
\begin{jolielisting}[][caption={Ad-hoc/External refactored API provider.},label={lst:ae:service}]
service PubCatWithAPIKey {
  outputPort pc { /* PubCat endpoint */ }(*\llabel{line:ae:outputport}*)
  inputPort ip { /* ... */ interfaces: PubCatInterfaceWithAPIKey }
  main { (*\llabel{line:ae:main-b}*)
    [ getAuthorPubs( request )( response ) {
      /* check validity of request.apiKey */
      if( isKeyValid ) {
        undef( request.apiKey ) /* remove API key before forwarding */ (*\llabel{line:ae:undef}*)
        getAuthorPubs@pc( request )( response ) /* forward call */
      } else { throw NotAuthorised( /* fault data */ ) }
    } ]
    [ getConfPubs( request )( response ) { /* as for getAuthorPubs */ } ]
  }(*\llabel{line:ae:main-e}*)}
\end{jolielisting}

\paragraph{Ad-hoc/Adjacent} Jolie supports running separate services in the same application with its native \jo{embedded} primitive. Thus, this strategy closely resembles the previous one, with the only difference being the deployment configuration of the two services \jo{PubCat} and \jo{PubCatWithAPIKey}.
First, the service \jo{PubCat} is promoted to an in-memory service by changing its location to \jo{"local"} (\cref{lst:o:service}, \cref{line:o:location}).
Then, we make the refactored API provider, \jo{PubCatWithAPIKey}, embed the original \jo{PubCat}.
This is achieved by the statement \jo{embed PubCat as pc} (\cref{line:aa:embed}), which instructs the Jolie runtime to load the service \jo{PubCat} alongside \jo{PubCatWithAPIKey} and make it reachable via an in-memory channel through the output port \jo{pc}.
These linguistic features allow for easily switching Jolie codebases between the Adjacent and External columns of the \EMI{} framework, changing the deployment strategy based on performance considerations (\ie, trade network overhead for CPU and memory consumption).
However, in this strategy, the two interfaces \jo{PubCatInterface} and \jo{PubCatInterfaceWithAPIKey} are still separate definitions that need to be manually kept in sync.
\begin{jolielisting}[][caption={Ad-hoc/Adjacent refactored API provider.},label=lst:aa:service]
service PubCatWithAPIKey {
  embed PubCat as pc(*\llabel{line:aa:embed}*)
  inputPort ip { /* (*\color{sgreen}Same as in \cref{lst:o:service}*) */ }
  main { /* (*\color{sgreen}Same as in \cref{lst:ae:service}*) */ }
}
\end{jolielisting}

\paragraph{Parametric/Adjacent}
To eliminate the coupling between refactored and original API providers, we leverage the Jolie language construct of an \emph{interface extender}, which uniformly extends the types of all operations in an API.
The extender \jo{APIKeyExtender} defined in \cref{lst:pa:service} adds the \jo{apiKey} field to all (\jo{*}) request messages and \jo{NotAuthorised} as a new potential faulty response.
\jo{APIKeyExtender} precisely describes the changes we have to apply to \jo{PubCatInterface} in order to obtain \jo{PubCatInterfaceWithAPIKey}.

\begin{jolielisting}[][caption={Parametric/Adjacent refactored API provider.},label=lst:pa:service]
interface extender APIKeyExtender {
  RequestResponse: *( {apiKey:string} )( void ) throws NotAuthorised
}

service PubCatWithAPIKey {(*\llabel{line:pa:service-b}*)
  embed PubCat as pc(*\llabel{line:pa:outputport}*)
  inputPort ip { /* ... */ aggregates: pc with APIKeyExtender(*\llabel{line:pa:aggregates}*) }
	courier ip { (*\llabel{line:pa:courier-b}*)
		[ interface PubCatInterface( request )( response ) {
      /* check validity of request.apiKey */
      if( isKeyValid ) { forward( request )( response ) }
      else { throw NotAuthorised( /* fault data */ ) }
		} ]
	} (*\llabel{line:pa:courier-e}*)
}(*\llabel{line:pa:service-e}*)
\end{jolielisting}

We use the interface extender to define the refactored API provider, service \jo{PubCatWithAPIKey} (\crefrange*{line:pa:service-b}{line:pa:service-e}).
The service embeds \jo{PubCat} (\cref{line:pa:outputport}) and refers to it through output port \jo{pc} as in \cref{lst:aa:service} above.
Differently, however, input port \jo{ip} now \jo{aggregates} \jo{pc} \jo{with} \jo{APIKeyExtender} (\cref{line:pa:aggregates}), which instructs Jolie to forward messages for the API of \jo{pc}, extended with \jo{APIKeyExtender}, to \jo{pc}.

Messages forwarded by means of aggregation (applications of \jo{aggregates}) can be intercepted by means of a \jo{courier} block.
A courier is a piece of code attached to an input port, which gets executed whenever one of the input port's operations is invoked.
The courier at \crefrange*{line:pa:courier-b}{line:pa:courier-e} implements the \apipattern{API Key} pattern for all operations of the interface \jo{PubCatInterface}.
Unlike a regular input choice, a courier can be parametric over the operation names of an interface:
\begin{equation*}
\mbox{\jo![ interface PubCatInterface( request )( response )\{ $\; B \;$ \} ]!}
\end{equation*}
where $B$ is the code that is executed on each invocation of an operation of \jo{PubCatInterface} on input port \jo{ip}, and which can then decide whether to \jo{forward} the request to the \jo{PubCat} service, or return the error message \jo{NotAuthorised}.
The \jo{forward} primitive automatically removes fields added by any interface extenders, so messages to \jo{pc} are well-typed.

\paragraph{Parametric/External} This solution differs from the previous one only on the deployment configuration of the refactored and original API providers, each having their own remote endpoint. The output port of \cref{lst:pa:service} must now describe the remote API endpoint of \jo{PubCat}.
\begin{jolielisting}
service PubCatWithAPIKey {
  outputPort pc { /* PubCat endpoint */ }
  inputPort ip { /* (*\color{sgreen}Same as in \cref{lst:pa:service}*) */ }
	courier ip { /* (*\color{sgreen}Same as in \cref{lst:pa:service}*) */ }
}
\end{jolielisting}

\paragraph{Parametric/Internal}
This strategy can be easily expressed in Jolie's syntax by adding an interface extender and courier to the original API provider.
\begin{jolielisting}
service PubCat {
  inputPort ip { /* ... */ interfaces: PubCatInterface with APIKeyExtender }
  courier ip { /* (*\color{sgreen}Same as in \cref{lst:pa:service}*) */ }
  main { /* (*\color{sgreen}Same as in \cref{lst:o:service}*) */ }
}
\end{jolielisting}
This is the only strategy that we could not test in Jolie, since its current interpreter does not support applying extenders to an interface that is not aggregated. This limitation is an implementation detail. Our study motivates the inclusion of this feature in future versions.

\subsection{Other patterns: \apipattern{Merge Endpoints} and \apipattern{Version Identifier}}
\label{sec:other-patterns}

We now illustrate how to introduce two other patterns: \apipattern{Merge Endpoints} and \apipattern{Version Identifier}. Differently from \apipattern{API Key}, these patterns are fully architectural, in the sense that they do not introduce behavioural changes but rather affect only how APIs are accessed.
We apply the Parametric/External strategy for both cases.

\apipattern{Merge Endpoints} exposes the operations of two endpoints through a single endpoint.
Suppose, for example, that we have a \jo{PubCat} service for a publication catalogue and a \jo{CitInd} service for citation indexing. We develop a new service, \jo{PublicationIndex}, that merges their APIs by using aggregation.
\begin{jolielisting}
service PublicationIndex {
  outputPort pc { // publication catalogue
    location: /* ... */ protocol: /* ... */
    interfaces: PubCatInterface
  }
  outputPort ci { // citation index
    location: /* ... */ protocol: /* ... */
    interfaces: CitIndInterface
  }
  inputPort ip {
    location: /* ... */ protocol: /* ... */
		aggregates: pc, ic
  } }
\end{jolielisting}
Note that aggregation requires the operations of the aggregated ports to have distinct names, which is in line with the pattern here. If this is not the case, one can use the other Jolie feature of redirection, explained in the next case.

\apipattern{Version Identifier} exposes two (or more) different versions of the same API under a single endpoint.
Here aggregation does not work, because the operation names in two versions of the same API likely overlap.
Jolie solves this problem by offering the APIs under different \emph{resource paths} at the same physical endpoint.
In the next example, input port \jo{ip} offers \jo{PubCatInterfaceV1} under path \jo{v1} and \jo{PubCatInterfaceV2} under path \jo{v2}.
Assuming that a client reaches the refactored API provider at location \jo{pubcat.com}, this means that version $1$ will be accessible at location \jo{pubcat.com/v1} and version $2$ at location \jo{pubcat.com/v2}.
\begin{jolielisting}
service PubCatWithAPIKey {
  outputPort pcv1 {
    location: /* ... */ protocol: /* ... */
    interfaces: PubCatInterfaceV1
  }
  outputPort pcv2 {
    location: /* ... */ protocol: /* ... */
    interfaces: PubCatInterfaceV2
  }
  inputPort ip {
    location: /* ... */ protocol: /* ... */
		redirects: v1 => pc1, v2 => pc2
  }
}
\end{jolielisting}
This approach does not alter the original (versions of) the APIs, by distinguishing between versions based on the accessed location. Therefore, clients just need to be connected to the right location.
An alternative to this approach is to extend the request types of all operations with a version identifier. However, this would require updating the clients to include this information.
Furthermore, response types would become less precise, since they would need to accommodate the possible responses across all versions.

\section{API Refactoring Recipes}
\label{sec:recipes}

In this section, we illustrate how our framework can be used to distill recipes that can be followed mechanically by programmers to apply an API refactoring. We cover the cases of a parametric implementation of the \apipattern{API Key} pattern and an ad-hoc implementation of the \apipattern{Pagination} pattern. The latter is representative of situations where obtaining efficiency requires big sacrifices in maintainability and isolation.

\newcommand{\recipetitle}[1]{\noindent\textbf{Refactoring recipe: #1}}
\newcommand{\recipepar}[1]{\vspace{.5em}\noindent\textit{#1}}

We start with our recipe for \apipattern{API Key}.

\begin{recipe}

\recipetitle{Introduce \apipattern{API Key} (Parametric)}

\recipepar{Intent.}
Introduce the \apipattern{API Key} pattern by means of a dedicated service that is parametric on the original API.

\recipepar{Participants and Preconditions.}
\begin{enumerate}
\item Participant: A Jolie \concept{service}, say \conceptname{Original}, exposing the API subject to refactoring as an interface, say \conceptname{OriginalAPI}.
\item Precondition: \conceptname{Original} offers \conceptname{OriginalAPI} through an \concept{input port} \conceptname{Original\-Input\-Port}.
\end{enumerate}

\recipepar{Refactoring steps.}
\begin{enumerate}
\item Introduce an \concept{interface extender} \conceptname{APIKeyExtender} that:
\begin{enumerate}
\item Extends the request message with a field \conceptname{apiKey} holding an \conceptname{API Key}.
\item Adds a faulty response message \conceptname{NotAuthorised}.
\end{enumerate}
\item Introduce a new \concept{service} \conceptname{OriginalWithAPIKey}:
\begin{enumerate}
\item Introduce a new \concept{output port} \conceptname{original}.
\item Choose between:
\begin{description}
\item[Choice 1 (External):]
Configure output port \conceptname{original} (at \conceptname{Original\-With\-API\-Key}) and input port \conceptname{OriginalInputPort} (at \conceptname{Original}) so that they communicate via the network.
\item[Choice 2 (Adjacent):]
Configure output port \conceptname{original} (at \conceptname{Original\-With\-API\-Key}) and input port \conceptname{OriginalInputPort} (at \conceptname{Original}) so that they communicate via local memory.
\end{description}
\item Introduce an \concept{input port} \conceptname{ip} that aggregates the output port \conceptname{original} and extends it with \conceptname{APIKeyExtender}.
\item Introduce a \concept{courier} for \conceptname{ip} that intercepts all operations of \conceptname{OriginalAPI} and:
\begin{enumerate}
\item Checks the validity of the \conceptname{API Key}.
\item If the key is valid, forwards the request to \conceptname{original}.
\item Otherwise, if the key is invalid, replies with the \conceptname{NotAuthorised} response.
\end{enumerate}
\end{enumerate}
\end{enumerate}

\recipepar{Postconditions.}
\begin{enumerate}
\item Invoking any operation \conceptname{op} at \conceptname{OriginalWithAPIKey} with a valid \conceptname{API Key} results into the same \conceptname{response} message as the invocation to \conceptname{op} at \conceptname{Original} without an \conceptname{API Key}.
\item Invoking any operation \conceptname{op} at service \conceptname{OriginalWithAPIKey} with an invalid \conceptname{API Key} results into an \conceptname{NotAuthorised} message.
\item Service \conceptname{OriginalWithAPIKey} becomes the only client of service \conceptname{Original}.
\end{enumerate}

\recipepar{Discussion and \EMI{} scoring.}
This recipe yields a parametric implementation, giving maintainability score \M3.
Choice 1 introduces network overhead, giving \E1 and \I3, while Choice 2 does not, yielding \E2 and \I2.
We get the following possible \EMI{} scores:
\begin{description}
\item[Choice 1 (External):] \E1 \M3 \I3.
\item[Choice 2 (Adjacent):] \E2 \M3 \I2.
\end{description}
\end{recipe}

We now present a recipe for the \apipattern{Pagination} pattern.
\pattern{Pagination} allows clients to retrieve smaller portions (`pages') of large data sets. The aim is to improve network and memory utilisation; this also addresses the stability antipattern of providing responses of unbounded size~\cite{N07}.
There are four variants of this pattern, corresponding to four different ways of identifying the page that the client wants~\cite{ZSLZ17,ZSLZP2022}. Here, we implement the offset-based version.

\begin{recipe}\recipetitle{Introduce \apipattern{Pagination} (Ad-hoc/Internal)}

\recipepar{Intent.}
Introduce the offset-based \apipattern{Pagination} pattern for an operation.

\recipepar{Participants and Preconditions.}
\begin{enumerate}
\item Participant: A Jolie \concept{service}, say \conceptname{Original}, exposing the operation subject to refactoring, say \conceptname{op}, as part of an interface, say \conceptname{OriginalAPI}.
\item Precondition: \conceptname{op} is a retrieval operation whose response type contains an ordered collection of items to be paginated.
\end{enumerate}
\recipepar{Refactoring steps.}
\begin{enumerate}
\item Change the definition of \conceptname{op} in \conceptname{OriginalAPI} to:
\begin{enumerate}
\item Extend the \concept{request type} with metadata fields specifying the offset of the requested page, the \conceptname{limit} of items per page, and the \conceptname{sort-criterion}, if more than one order exists for items in the collection;
\item Extend the \concept{response type} with fields describing the
response page such as the page number \conceptname{offset}, items per page
\conceptname{limit}, \conceptname{sort-criterion}, and \conceptname{total} number
of pages.
\item Add a faulty response message \conceptname{InvalidPageRequest} in case of invalid page metadata.
\end{enumerate}
\item Change the implementation of \conceptname{op} to:
\begin{enumerate}
\item Validate the page metadata fields (and reply immediately with \conceptname{Invalid\-Page\-Request} in case of failure).
\item Paginate the requested data, possibly by leveraging features of the data\-base query language (like \conceptname{OFFSET} and \conceptname{LIMIT} for SQL).
\item Reply with the requested page and its metadata.
\end{enumerate}
\end{enumerate}
\recipepar{Postconditions.}
\begin{enumerate}
\item Calling \conceptname{op} to request page with a given
\conceptname{offset} and size \conceptname{limit} results into the
items of the \conceptname{collection} returned by the original \conceptname{op} from
position \conceptname{offset * limit} to position \conceptname{offset * limit + limit}.
\end{enumerate}

\recipepar{Discussion.}
Delegating the pagination to the query language of the database in use achieves efficiency score \E3.
However, since it also modifies the implementation of the specific operation, we obtain maintainability \M1 and isolation \I1.
The overall \EMI{} score is therefore \E3 \M1 \I1.

\recipepar{Considerations on alternative implementations.}
The design smells that motivate the introduction of the \apipattern{Pagination} pattern are about poor efficiency and thus the Ad-hoc/Internal strategy is a natural choice. 
If Ad-hoc/Internal is undesirable, other strategies can still be adopted at the cost of high decreases in efficiency.
The key problem is distribution. Choosing an Adjacent strategy would still imply that the original API provider fetches all data from its database, but at least this would be `cut' by the refactored API provider before it is sent back to clients.
The same holds for an External strategy, but in this case we would pay also the cost of network communication (of the whole data set) between the refactored and original API providers.
\end{recipe}

\section{Conclusion}
\label{sec:conclusion}
We have introduced the \EMI{} framework, the first conceptual framework for navigating the implementation aspects of API refactorings.
While broad and technology-agnostic, our scores are informative when it comes to key design decisions on the implementation of API patterns.

Our study opens up at least two interesting lines of future work.

First, in line with previous work~\cite{SZ23}, we have focused on presenting API refactorings that \emph{add} a pattern. However, our Adjacent and External strategies make it immediate to \emph{remove} a pattern later on. We think that enabling the modular activation and deactivation of patterns is an interesting direction.

The second line of future work deals with exploring additional aspects on top of efficiency, maintainability, and isolation. 
These aspects are in line with a previous survey on what qualities are important in practice, but there are also others that merit consideration, like scalability and usability~\cite{SZ21,BWZ2024}.
We think that scalability would be a first natural extension of our framework, as it is closely related to efficiency and isolation but not completely captured by them.

\subsubsection*{Acknowledgements}
We thank Sandra Greiner for the useful discussions and feedback on a draft of
this paper.

\bibliographystyle{splncs04}
\bibliography{bibliography}
\end{document}